\documentclass[aps,pre,showpacs,twocolumn]{revtex4}

\usepackage{amsmath}
\usepackage{epstopdf}
\usepackage{graphicx,textcomp}
\usepackage{dcolumn,amsmath,amsthm,amscd,amsfonts,amssymb,epsfig,graphics,graphicx,eucal}

\begin{document}


\title{Dissipative defect modes in periodic structures}

\author{Yaroslav V.  Kartashov$^1$, Vladimir V. Konotop$^2$, Victor A. Vysloukh$^3$, and Lluis Torner$^1$}

\address{$^1$ICFO-Institut de Ciencies Fotoniques, and Universitat Politecnica de Catalunya, Mediterranean Technology Park, 08860 Castelldefels (Barcelona), Spain
}
\address{
 $^2$Centro de F\'{\i}sica Te\'orica e Computacional and Departamento de F\'{\i}sica, Faculdade de Ci\^encias, Universidade de Lisboa,
Avenida Professor Gama
Pinto 2, Lisboa 1649-003, Portugal
}
\address{
$^3$Departamento de Fisica y Matematicas, Universidad de las Americas - Puebla, Santa Catarina Martir, 72820, Puebla, Mexico
}

\begin{abstract}
We show that periodic optical lattices imprinted in cubic
nonlinear media with strong two-photon absorption and localized
linear gain landscapes support stable dissipative defect modes in
both focusing and defocusing media. Their shapes and transverse
extent are determined by the propagation constant that belongs to
a gap of the lattice spectrum which, in turn, is determined by the
relation between gain and losses. One-hump and two-hump
dissipative defect modes are obtained.

\end{abstract}

\maketitle


\noindent

Implementation of stable spatial localization of light is a topic
of continuously renewed interest because  of its fundamental
importance and potential applications. The use of
structures with shallow periodic refractive index modulations
is a fruitful strategy to generate
and to control soliton-like light packets (see,
e.g.~\cite{PhysRep,KVT_review} for recent reviews). Localization, alias defect modes, 
can be achieved in linear periodic systems 
with structure defects.
 The interplay between nonlinearity, periodicity of the underlying
structure, and local lattice deformations results in appearance of
new types of localized modes and important
modification of stability of modes existing in linear regime (see
~\cite{Fedele,Chen,Efremidis,Fe} in the context of nonlinear
optics and in~\cite{BraKonPer1,BraKonPer2} in the context of
Bose-Einstein condensates). Further development of this concept
can be achieved in non-conservative systems where dissipation and
gain are present, in particular, using dissipative defects. The
corresponding idea, where the defect is produced by localized
gain, has been suggested upon study of interactions of gap
solitons in shallow fiber Bragg gratings~\cite{MMC} and in media
of two-level atoms~\cite{MA}, as well as for stabilization of
spatial dissipative solitons
 described by the
complex Ginzburg-Landau equation~\cite{malomed_delta}
(i.e. relevant to cavity solitons).
Our statement is also related to the studies of the discrete Ginzburg-Landau equation, which possesses unstable localized solutions~\cite{MAA},  one of the stabilization mechanisms for which is the quintic nonlinearity~\cite{EC}.

In this Letter we 
report existence and stability of
dissipative defect modes in periodic optical lattices (we
distinguish them from solitons since the modes are pinned by
several "defect"  channels where gain is realized and exist only
due to competition between the gain and nonlinear losses). Some of
the properties of such modes appear to be very different from
their conservative
counterparts~\cite{Fedele,Chen,Efremidis,Fe,BraKonPer1,BraKonPer2}.
Being attractors they possess enhanced stability and can be excited
from sufficiently large class of initial conditions.
  
Specifically, we consider propagation of laser radiation in a
lattice with spatially localized linear gain and nonlinear losses
(two-photon absorption) that can be described by the nonlinear
Schr\"odinger equation for the dimensionless light field amplitude
$q$:
\begin{eqnarray}
iq_\xi = - \frac 12 q_{\eta\eta} -R(\eta)q +i\gamma(\eta)q-\sigma |q|^2q - i\alpha |q|^2q
\label{GPEq}
\end{eqnarray}
Here $\eta$ and $\xi$  are the normalized transverse and
longitudinal coordinates, respectively, $\sigma=1$ ($\sigma=-1$)
for focusing (defocussing) media,  $\alpha>0$ characterizes
nonlinear losses,    $R(\eta)$ is a periodic function,
and $\gamma(\eta)$
describes the linear gain which is nonzero only in a finite
domain, i.e. $\gamma(\eta)\equiv 0$ for $|\eta|>\eta_0$ ($\eta_0$
is a positive constant).

One of our central results is that system (\ref{GPEq}) supports
stable localized modes in both focusing and defocusing nonlinear
media. The complex amplitude of such modes can be written in the
form $q(\xi,\eta) = e^{ib\xi+i\theta(\eta)} w(\eta)$, with a real
propagation constant $b$, real stationary amplitude $w$ and phase
$\theta$, and with the zero asymptotes: $  w(\eta)\to 0$ at
$\eta\to\pm \infty$. Consequently we rewrite (\ref{GPEq}) in the
form of equations for real-valued functions $w(\eta)$ and
$j(\eta)\equiv\theta_\eta w^2$:
 \begin{eqnarray}
\label{rho}
bw=\frac{w_{\eta\eta}}{2}-\frac{j^2}{2w^3}
+\sigma w^3+Rw,\quad
j_\eta
=2\gamma w^2-2\alpha w^4
\end{eqnarray}

\begin{figure}[h]
   \begin{center}
 \includegraphics[width=\columnwidth]{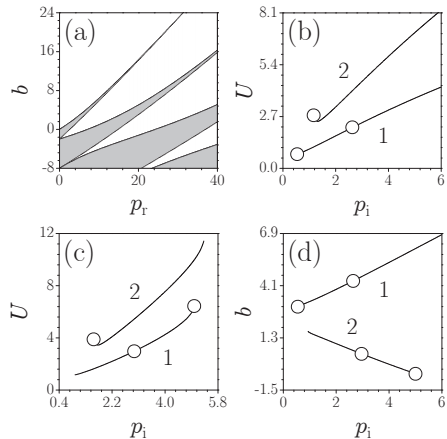}
  \end{center}
\caption{(a) The band spectrum of $R(\eta)$. For
$p_r=5$ the lower edge of the semi-infinite gap is at $b=2.875$
while the first finite gap corresponds to the interval
$(-0.645,1.840)$.  Shadowed (empty) regions indicate the allowed
bands (gaps). Dependence $U$ {\it vs} $p_i$ for the one-hump
(curve 1) and two-hump (curve 2) modes is shown in (b) for the
focusing   and in (c) for defocusing media.  (d) Propagation
constant {\it vs}   gain for one-hump modes in focusing (curve 1)
and defocusing (curve 2) media. Circles in panels (b)-(d)
correspond to the modes shown in Fig.~\ref{fig2}. In all the cases
$\alpha=1$.} \label{fig1}
\end{figure} 
The propagation constant of an exponentially localized mode must
belong to a gap in the spectrum of the periodic "potential"
$R(\eta)$,  i.e. to an empty domain in Fig.~\ref{fig1} (a).  This
is another important finding of the present work, which suggest a
way to classify defect modes. Indeed, let us consider the limit
$\eta\to \infty$. From the second of Eqs.~(\ref{rho}) it follows
that $j=2\alpha\int_{\eta}^{\infty}w^4d\eta$ for $\eta>\eta_0$.
Hence, (i) there exists an energy flow outwards the
impurity, $j>0$, and (ii) the following relation (obtained using
l'H\^opital's rule)
$
\lim_{\eta\to \infty} {j }/{w^2 }=- \alpha
\lim_{\eta\to \infty} { w^3 }/{w_\eta }
$
holds.
For localized modes with the finite energy flow
$U=\int_{-\infty}^{\infty}w^2(\eta)d\eta$, i.e. with $w^2$
decaying more rapidly than $1/\eta$, one obtains that the above
mentioned limit is zero as long as $w_\eta$ is sign-definite.
Otherwise, if $w_\eta$  changes sign at  large
$\eta$, one also verifies that the established condition $j>0$ can
hold only if the above mentioned limit is zero. Hence, $j^2/w^3$
decays faster than $w$ and can be neglected at  $\eta\to\infty$
and  $b$ satisfies the Hill equation
$
b\tilde{w}=\frac 12\tilde{w}_{\eta\eta}+R(\eta)\tilde{w}.
$
Thus, to guarantee the decaying asymptotic of 
the mode,  $b$ 
must belong to a gap of the lattice spectrum.  
We emphasize the importance of the sufficiently fast decay of the linear dissipation coefficient 
for the last conclusion.

In contrast to conservative defect modes, the stationary states in
our system exist not only due to balance between   nonlinearity
and refraction/diffraction of the lattice, but also due to balance
between  the linear gain and nonlinear losses ~\cite{Akhmediev}, 
which follows from Eq.~(\ref{GPEq}):
$
\int_{-\infty}^{\infty}\gamma(\eta)w^2d\eta=\alpha\int_{-\infty}^{\infty}w^4d\eta.
$
Therefore,
the propagation constant $b$  and the energy flow  $U$  are determined by linear gain $\gamma$  and by the nonlinear losses $\alpha$.

Turning to numerical study of the modes, we chose $R(\eta)=p_r\cos^2(\Omega  \eta)$ to model the lattice, whose spectrum is shown in Fig.~\ref{fig1}(a).
We consider localized linear gain profiles, i.e., we assume that
 $\gamma(\eta)\propto R(\eta)$ on  one or several
 periods of $R(\eta)$. For instance, if amplification is realized only in one period, then $\gamma(\eta)=p_i\cos^2(\Omega  \eta)$  for $|\eta|\leq\eta_0=\pi/(2\Omega)$ and $\gamma\equiv 0$ for $|\eta|>\eta_0$. Notice, that other functional profiles of the gain in amplifying lattice channels yield qualitatively similar results.
  Further we set $\Omega=2$, $p_r=5$ and vary $p_i$ and
 $\alpha$.

\begin{figure}[h]
  \begin{center}
   \includegraphics[width=\columnwidth]{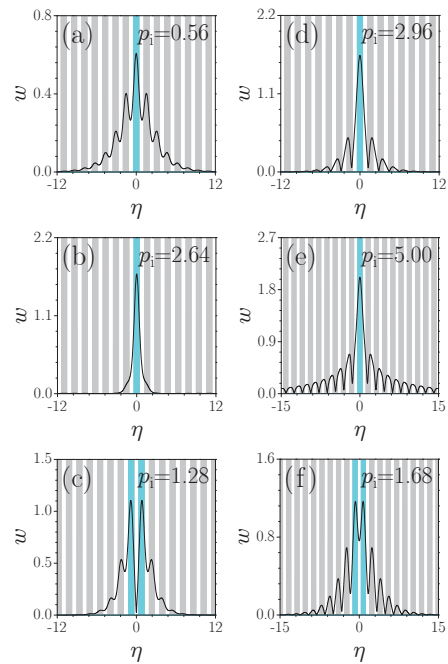}
    \end{center}
\vspace{-1cm}
\caption{Profiles of one-hump (one-channel gain) (a),(b), (d), (e)  and two-hump (two-channels gain) (c), (f) modes in focusing (left column) and defocusing (right column) media. The humps in two-hump modes in focusing and defocusing media are out-of-phase and in-phase respectively.
}
\label{fig2}
\end{figure}

When the propagation constant belongs to the semi-infinite gap,
the defect modes can be obtained in focusing medium only. The
emerging branches of the modes  are shown in Fig.~\ref{fig1} (b),
while examples of their profiles are presented in the left column
of Fig.~\ref{fig2}.
As we mentioned, $b$ is determined by the gain/loss as shown in
Fig.~\ref{fig1}(d). Notice that depending on whether the gain is
applied to one or two channels the attractor is a one-hump [panels
2(a),2(b)] or a two-hump [panel 2(c)] mode.

For $b$ in the first finite gap we have found the stable defect modes in the defocusing medium. Their branches are shown in  Fig.~\ref{fig1} (c), and the examples of the amplitude configurations are given in the right-column panels of Fig.~\ref{fig2}.
 
Like in the conservative case, when the propagation constant approaches a gap edge defect modes extend far beyond the lattice channels where amplification is applied, exhibiting pronounced shape oscillations which approaches properly normalized Bloch states $\tilde{w}$ bordering the respective gap edge.
At fixed $\alpha$  in a focusing medium an increase of the gain $p_i$ results in progressive increase of the mode peak intensity and in gradual contraction of light into the amplifying channel [Fig.~\ref{fig2}(a),  (b)].
 In defocusing medium the peak amplitude  also grows with $p_i$
but the modes possess strongest localization at intermediate values of $p_i$ and may extend dramatically across the lattice when the gain intensity becomes too high [Fig.~\ref{fig2}(d),(e)]. The energy flow $U$ monotonically grows with $p_i$  [Fig.~\ref{fig1}(b), (c)].
For sufficiently small and moderate nonlinear losses one-hump modes exist even at $p_i\to 0$  in both focusing and defocusing media. In this limit  $b$  approaches the band-edge, the amplitude of the mode becomes small and its width grows dramatically. The nonlinear losses diminish as well, and  the balance between gain and losses requires  the linear gain to  go to zero. 
Notice, however, that for sufficiently high nonlinear losses stable defect modes can be found only above certain critical value of $p_i$ and in this parameter range stable modes exist above threshold energy flow.
 
Higher-order modes  possess energy flow threshold for their existence  
[curves 2 in Fig.~\ref{fig1}(b), (c); we confirmed this for all modes in focusing (defocusing) medium that change (do not change) their sign between neighboring lattice sites]. 
Increase of  $p_i$ results in monotonic growth of $b$  in focusing medium accompanying mode contraction [Fig.~\ref{fig1}(d), curve 1] and decrease of $b$ in defocusing medium resulting in mode expansion across the lattice when $b$  reaches the lower edge of the gap at sufficiently large  $p_i$ [Fig.~\ref{fig1}(d), curve 2].

\begin{figure}[h]
\begin{center}
   \includegraphics[width=\columnwidth]{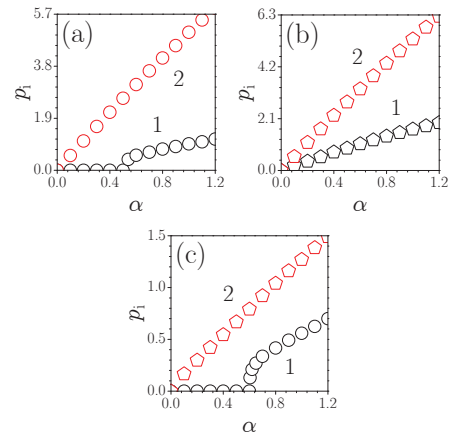} 
   \end{center}
   \vspace{-1cm}
\caption{Domains of existence for stable one-hump and two-hump modes (boundaries denoted by circles and pentagons, respectively) in defocusing   (a), (b) and focusing (c) media. In the defocusing medium either one-hump (a) or two-hump (b) modes modes exist between the curves 1 and 2. In the focusing medium one-hump and two-hump modes exist above the curves 1 and 2, respectively.}
\label{fig4}
\end{figure}

Representative examples of the field modulus distribution of defect modes
are shown in Fig.~\ref{fig2}. As it was mentioned above, when amplification is realized only in one lattice channel one obtains simplest one-hump modes [Figs.~\ref{fig2}(a),(b)]. When two central channels are amplifying we found  two-hump modes in focusing [Fig.~\ref{fig2}(c)] and in defocusing [Fig.~\ref{fig2}(f)] media. These states were obtained by direct simulation of Eq.~(\ref{GPEq}) with different inputs (ranging from localized input beams to extended noisy distributions) up to a huge distance. This guarantees that final states that do not change upon propagation are  stationary attractors of
Eq.~(\ref{GPEq}) [solutions of Eq.~(\ref{GPEq}) with other symmetries are possible too, but they are unstable and not shown here]. Notice that for larger number of amplifying channels  more complicated modes with larger number of humps can be obtained.

The domains of existence of dissipative lattice modes on the plane $(\alpha,p_i)$  are shown in Fig.~\ref{fig4}. In the focusing medium one-hump modes exist for all values of $p_i$ above curve 1 in Fig.~\ref{fig4}(c)
while two-hump modes are found for  $p_i$ values above curve 2 (the minimal value of  $p_i$ necessary for existence of two-hump modes grows almost linearly with $\alpha$). The situation is more complicated in defocusing medium where one-hump [Fig.~\ref{fig4}(a)] and two-hump [Fig.~\ref{fig4}(b)] modes exist only in a finite domain of $p_i$ between curves 1 and 2. The presence of upper limit for the linear gain is connected with delocalization of high-power modes whose propagation constants approach lower edge of first finite gap with increase of $p_i$. The domain of existence of two-hump modes in terms of $p_i$  values expands monotonically with increase of $\alpha$  [Fig.~\ref{fig4}(b)].

Summarizing, we showed that stable defect modes can exist in
optical lattices imprinted in nonlinear medium with strong
two-photon absorption provided that gain is realized in one or
several lattice channels.

\end{document}